%% file: main.tex
\documentclass[sigconf]{acmart}
\AtBeginDocument{%
  }
    
\usepackage{packages/packages}
\usepackage{packages/macros}

\setcopyright{acmlicensed}
\copyrightyear{2018}
\acmYear{2018}
\acmDOI{XXXXXXX.XXXXXXX}
\acmConference[Conference acronym 'XX]{Make sure to enter the correct
  conference title from your rights confirmation email}{June 03--05,
  2018}{Woodstock, NY}
\acmISBN{978-1-4503-XXXX-X/2018/06}

\begin{document}

\title{\ours: Co-Optimizing RAG Quality and Performance \\ From the Vector Database Perspective}  

\author{Wenqi Jiang}
\affiliation{%
  \institution{Systems Group, ETH Zurich}
  \country{}
}
\email{wenqi.jiang@inf.ethz.ch}


\renewcommand{\shortauthors}{Wenqi Jiang et al.}

\sloppy
\input{sections_submission/abstract}
\settopmatter{printacmref=false}
\renewcommand\footnotetextcopyrightpermission[1]{}
\maketitle

\input{sections_submission/intro}
\input{sections_submission/rag-complexity}
\input{sections_submission/goal}

\input{sections_submission/solution}
\input{sections_submission/conclusion}


\bibliographystyle{ACM-Reference-Format}
\bibliography{ref}

\end{document}

%% file: sections_submission/abstract.tex
\begin{abstract}

Retrieval-augmented generation (RAG) has emerged as one of the most prominent applications of vector databases.
By integrating documents retrieved from a database into the prompt of a large language model (LLM), RAG enables more reliable and informative content generation.
While there has been extensive research on vector databases, many open research problems remain once they are considered in the wider context of end-to-end RAG pipelines.
One practical yet challenging problem is how to jointly optimize both system \textit{performance} and generation \textit{quality} in RAG, which is significantly more complex than it appears due to the numerous knobs on both the algorithmic side (spanning models and databases) and the systems side (from software to hardware).
%
%
%
In this paper, we present \ours, a three-pillar blueprint for quality–performance co-optimization in RAG systems. \ours comprises:
(1) \ir, an intermediate representation that serves as an abstraction layer to decouple quality and performance aspects;
(2) \perf, a cost model for estimating system performance given an \ir; and
(3) \plan, a plan exploration algorithm that searches for high-quality, high-performance RAG configurations.
We believe this three-pillar blueprint will become the de facto paradigm for RAG quality–performance co-optimization in the years to come.


%
%

\end{abstract}

%% file: sections_submission/intro.tex
\section{Introduction}
\label{sec:intro}





\begin{figure}[t]
  \centering
  \includegraphics[width=\linewidth]{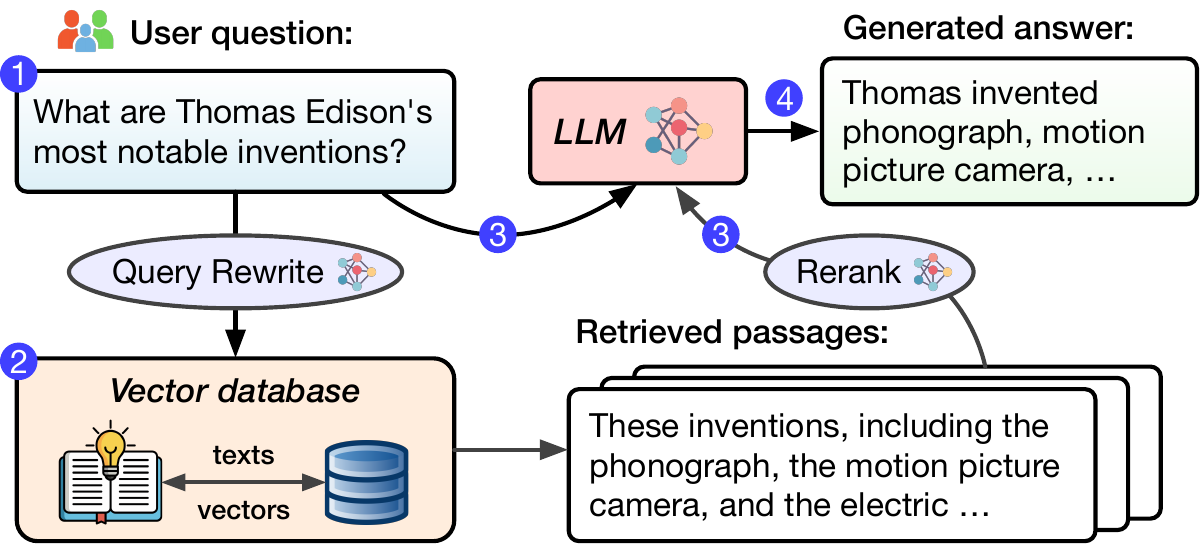}
  \vspace{-1.5em}
  \caption{A simple RAG pipeline with models and a database.}
  \vspace{-1em}
  \label{fig:rag-example}
\end{figure}

Vector database has become a hot topic in recent years due to the rapid advancement of machine learning and generative AI.
Among its many applications, Retrieval-Augmented Generation (RAG) has emerged as one of the most prominent use cases in the LLM era.
By incorporating relevant documents retrieved from databases into the LLM prompt as shown in Figure~\ref{fig:rag-example}, RAG improves the reliability, timeliness, and comprehensiveness of generated responses~\cite{borgeaud2022improving, lewis2020pre, edge2024local}.
As a result, RAG has become the industry standard for knowledge-intensive natural language processing tasks~\cite{borgeaud2022improving, shao2024scaling, wang2023instructretro, lazaridou2021mind, lewis2020pre}.


\textbf{Well-Tuned Vector Databases Alone Are Not Enough for RAG Serving.}
We have witnessed extensive research on optimizing query processing in vector databases (referred to as retrievals throughout this paper for brevity), particularly in terms of quality (e.g., recall) and performance (e.g., QPS, latency), often on standard benchmarks with a predefined dataset including a fixed number of encoded vectors, a specified Top-$K$, and a target recall requirement.
While research on retrieval itself is undoubtedly fundamental, many open problems remain when integrating vector databases into end-to-end RAG pipelines, because the previous assumptions (optimizing retrievals in isolation on well-defined benchmarks) no longer hold.
\textit{On the other hand, there is a vast number of configurable knobs in RAG spanning both algorithmic and system aspects, all of which influence end-to-end RAG system \textit{performance} (e.g., latency, throughput) and generation \textit{quality} (e.g., accuracy, reliability), making it essential to move beyond tuning vector databases in isolation.}

\textbf{Algorithm Configuration Space.}
There are numerous algorithmic factors that influence the generation quality of a RAG pipeline. 
For example, how is the database configured? This includes decisions such as document chunking strategies, choice of embedding models, index configurations, and retrieval quality requirements such as recall. 
In addition to the database, should auxiliary models be used before and after retrieval? These may include a query rewriter that rephrases or decomposes the user’s original question, and a result reranker that re-scores the relevance of the retrieved Top-$K$ documents.
%
Once relevant content is retrieved, how many documents should be incorporated into the prompt? 
%
%
Many factors that are typically taken for granted in standard retrieval benchmarks become essential variables in this end-to-end context.
Notably, the optimal configuration not only depends on these many factors, but also (a) varies across applications and coupled LLMs, and (b) evolves as new retrieval techniques and models emerge.

\textbf{System Design Space.}
There is also a wide range of system design choices beyond the database itself, as performance and efficiency must be optimized for the entire RAG pipeline as a whole.
First, there are software-level options, such as the choice of databases and LLM serving frameworks.
Second, there are various hardware options: depending on the type and availability of inference hardware, should the database run on CPUs (in-memory or on-disk), leverage accelerators such as GPUs or FPGAs, or combine different hardware in a heterogeneous setup?
%
Third, end-to-end RAG scheduling and resource management decisions must be made, such as how to allocate hardware resources across RAG components (e.g., database, query rewriter, result reranker, and the main LLM) and how to configure batching policies.
Just as optimal RAG algorithm configurations evolve, optimal system configurations for a given RAG algorithm will also shift over time as software and hardware continue to advance.



\textbf{Research Gap: Co-Optimizing Quality and Performance.} 
Given the many additional dimensions introduced by RAG, the optimization objective naturally shifts from optimizing recall and QPS in standalone database to \textit{advancing the performance–quality Pareto frontier for end-to-end RAG serving.}
Due to the complexity of the RAG landscape as explained above, it is not surprising that existing research has focused on optimizing either the quality or performance.
%
%
One might suggest that performance–quality co-optimization for RAG can be achieved through expert-based tuning, but such an approach is often suboptimal and, in many cases, impractical.
This is not only because it is challenging to find an individual or a tightly coordinated team with expertise spanning RAG algorithms, database systems, and inference software/hardware.
More importantly, as discussed earlier, the optimal RAG algorithm–system configuration varies across applications and evolves over time with the emergence of new retrieval techniques, models, and software/hardware stacks, making expert knowledge brittle and quickly outdated.

%
%

\begin{figure}[t]
  \centering
  \includegraphics[width=\linewidth]{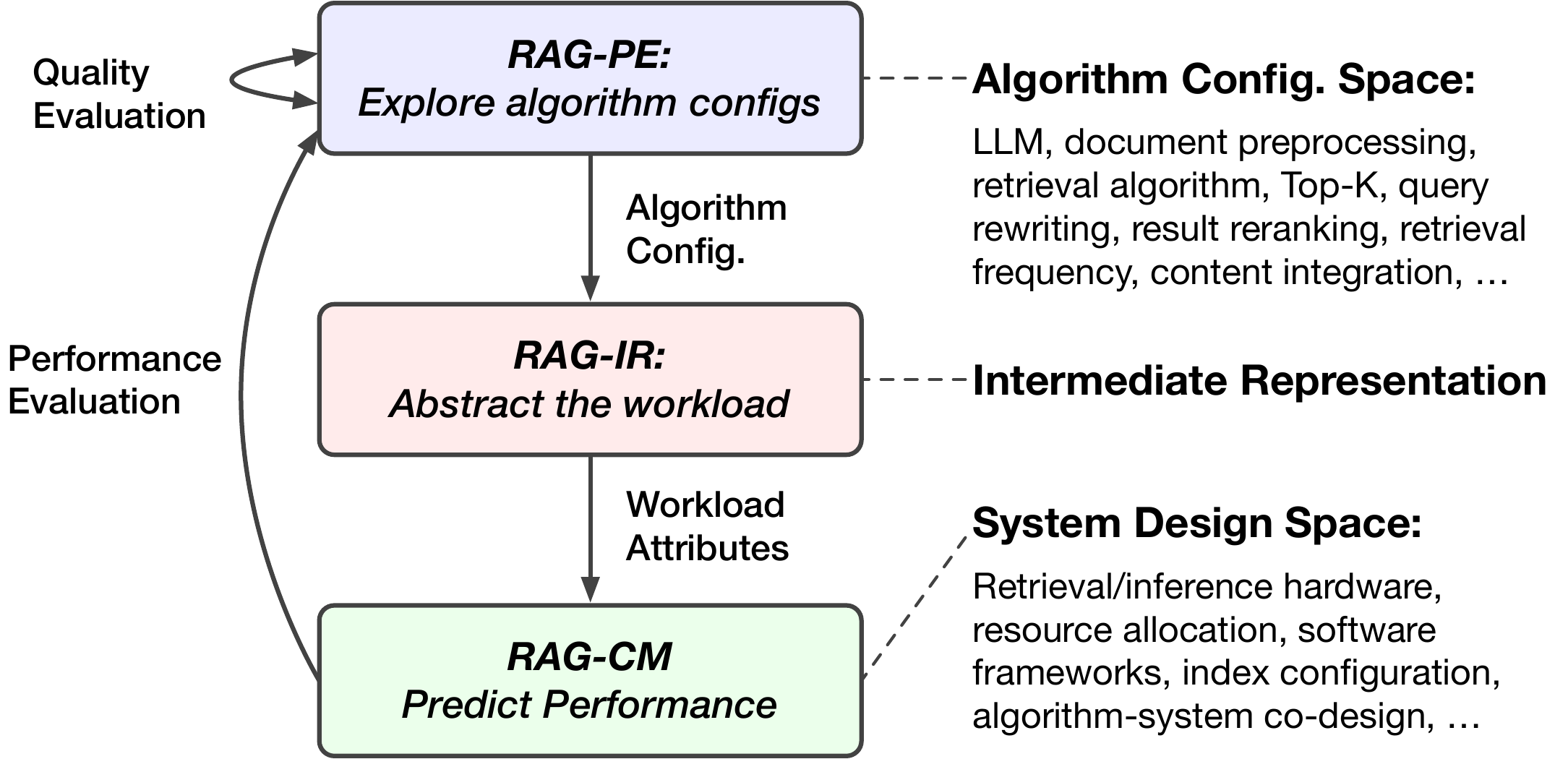}
  \vspace{-1.5em}
  \caption{\ours systematically explores the performance–quality Pareto frontier for RAG serving.}
  \vspace{-1.5em}
  \label{fig:overview}
\end{figure}

\textbf{\ours Overview.}
\textit{In this paper, we propose \ours, a three-pillar blueprint for systematically co-optimizing quality and performance in RAG serving.}
As shown in \Cref{fig:overview}, \ours consists of: (1)~\ir, an \underline{I}ntermediate \underline{R}epresentation and abstraction layer that bridges RAG algorithms and systems; (2)~\perf, a \underline{C}ost \underline{M}odel on the systems side that predicts performance given a RAG algorithm; and (3)~\plan, a \underline{P}lan \underline{E}xploration strategy that navigates the RAG configuration space. 
The key idea is to let \plan iteratively explore the RAG algorithm space, where in each iteration, the generation quality is evaluated and the maximum achievable system performance is predicted by \perf using the algorithm's \ir representation.
Guided by prior evaluations of quality and performance, \plan predicts the next configuration to explore that is likely to approach the performance–quality Pareto frontier.
We now elaborate on each of these three pillars:


\textbf{Abstraction.}
First, \ir provides a well-defined abstraction between the quality and performance dimensions, while remaining extensible to emerging RAG algorithm–system co-design techniques.  
The goal of this abstraction is to allow ML and systems researchers to consider the quality and performance optimization separately.
At its core, \ir is a collection of dataflow graphs, where each node represents a model or a database component, annotated with detailed attributes such as model size, sequence length, and search parameters. Edges between nodes capture the volume of data exchanged between components.  
This abstraction enables a clean separation between properties that impact only quality (e.g., the version of the model used with the same architecture and size) and those that influence only performance (e.g., the choice of hardware for retrieval).
As algorithm-system co-design techniques that may affect quality continue to emerge --- such as KV-cache reuse~\cite{yao2024cacheblend} or speculative retrieval~\cite{jiang2025piperag} --- \ir allows such techniques to be explicitly annotated within the pipeline as well.

\textbf{Cost Model.}
The second component of \ours \perf, a cost model that estimates system performance given a specific RAG algorithm configuration.  
In particular, the algorithm configuration is first turned into its corresponding \ir representation. 
Based on the available hardware and software stack, \perf then returns either the actual performance $p$ or an estimated performance $\hat{p}$.  
\perf can take various forms, including ML-based cost models, analytical models, or empirical lookup tables derived from profiling on real hardware.  
\perf enables: (1) algorithm researchers to explore quality–performance trade-offs without needing to understand the underlying systems; and (2) systematic exploration of the quality–performance Pareto frontier, as we will discuss next.

%

\textbf{Plan Exploration.}
The third pillar of \ours is \plan, a plan exploration algorithm that iteratively navigates the RAG algorithm configuration space to identify the (near-) Pareto frontier between quality and performance, while minimizing the number of evaluations.
In each iteration, \plan evaluates the system performance (via \perf and \ir) and generation quality of a given algorithm configuration, then predicts the next configuration to explore based on previous quality–performance observations.
Various exploration strategies can be used within \plan, such as grid search, genetic algorithms, or reinforcement learning.  
The search space can also be constrained based on deployment requirements, such as restricting the set of available models or targeting a specific hardware cluster.
%
%


\textbf{Research Opportunities.}
By proposing these three pillars, we also discuss three key research gaps.
First, should index configuration be part of the RAG workload abstraction?
While it may seem natural for algorithm researchers to specify a recall requirement and leave system optimization to infrastructure experts, two indexes achieving the same recall can return different document sets for individual queries, potentially leading to different RAG generation quality.
Second, there is a clear need for more general and accurate cost models for retrievals. These models should support a variety of index types, distributed retrieval, and different hardware backends.
Finally, predicting the next database configuration to evaluate in the vast RAG configuration space is crucial, as rebuilding the database can be extremely costly ---for example, switching to a new index type, modifying the chunking strategy, or re-encoding vectors with a different embedding model.

\textbf{Contributions.}  
To our knowledge, \ours is the first blueprint for systematic quality–performance co-optimization in RAG serving.  
We not only introduce the workload abstraction (\ir), the cost model (\perf), and the plan exploration algorithm (\plan), but also outline future research opportunities in each area.
We believe \ours will become a de facto paradigm for optimizing RAG serving and a foundation for continued research in this space.



%



%% file: sections_submission/rag-complexity.tex
\section{RAG Complexity}
\label{sec:complexity}

\input{tables/algorithm_space}
\input{tables/system_space}

While the key idea of RAG --- combining an LLM with a database --- may seem simple , RAG serving is significantly more complex than it appears due to the vast algorithmic configuration space and the many system-level deployment options.



\subsection{Algorithm Configuration Space}
\label{sec:complexity:algorithm}

We define a RAG algorithm configuration as a combination of model selection, database configuration, and the method used to integrate retrieved content from the database into the model --- all these factors influence generation quality, and the optimal configuration varies across applications (medical question answering, general-purpose RAG, etc.). We summarize the detailed configurations in Table~\ref{tab:rag-config-space} and briefly explain the key ones below.

From the database perspective, there are many configurations to be tuned.  
For example: How many documents are indexed? Are documents chunked into shorter passages? Should we use a vector-based index or a keyword-based one? Which embedding model is used to convert documents into vectors? What is the size of Top-$K$ nearest neighbors retrieved per query? What level of retrieval quality is required (e.g., 95\% recall)? Should the natural language query be rewritten before retrieval? Should retrieval results be reranked or filtered after retrieval?  

From the database-model interaction perspective, there are not only choices of which LLM to use, but also different ways to integrate the retrieved content.  
For example: Is retrieval performed once at the beginning of generation, or multiple times during generation? How is the retrieved content incorporated --- via prompt tokens or through KV-cache insertion (without cross-attention mechanism between documents)? Is the model guided to ignore retrieved documents that are less relevant?

Note that for a given application, there may be multiple RAG algorithm configurations that can achieve the same generation quality.  
For example, to ensure sufficient knowledge is integrated into the LLM output, one could choose among the following strategies:  
(a) build a high-quality index and enforce a high-recall retrieval policy, then pass only a small number of top-ranked documents to the LLM~\cite{borgeaud2022improving};  
(b) use a relaxed recall constraint during retrieval, return a larger set of candidate documents, and apply a reranker model to select the most relevant ones~\cite{glass2022re2g}; or  
(c) directly feed a large number of retrieved documents into the LLM and let the model implicitly rank and filter them during generation~\cite{asai2023self}.  
While these different algorithmic choices may yield comparable quality, they can result in different performance characteristics in terms of latency, throughput, and the distribution of workload across RAG components.

\subsection{System Design Space}
\label{sec:complexity:system}

Given a RAG algorithm configuration, there are many ways to implement a RAG system, as overviewed in Table~\ref{tab:rag-system-design-space}. 
From the hardware perspective, we need to choose the type and amount of compute resources (e.g., CPUs and/or accelerators), and map each RAG component (including models and databases) to the allocated hardware resources.  
From the software perspective, we must not only choose inference libraries and vector databases, but also choose the indexing algorithm and its parameters.

Apart from these basic system configurations, there are emerging cross-stack performance optimizations that modify the original RAG algorithm and can consequently affect generation quality.
One example is storing and reusing per-document KV caches for frequently accessed documents~\cite{yao2024cacheblend, jin2024ragcache}, allowing the model to directly integrate the stored KV cache of retrieved documents without recomputation. While this method reduces compute cost, it discards cross-attention between the Top-$K$ retrieved documents, which may degrade output quality~\cite{yao2024cacheblend}.  
Another example is speculative retrieval, which hides retrieval latency during generation by prefetching documents using approximate queries --- especially useful in iterative retrieval scenarios~\cite{jiang2025piperag, zhang2024accelerating}. While this technique improves both latency and throughput, the approximation may result in suboptimal retrieved content. 
In summary, these cross-stack algorithm-system co-design solutions offer performance gains but may introduce trade-offs in quality, and should be selectively enabled based on application requirements.




%% file: tables/algorithm_space.tex
\begin{table*}[t]
\centering
\caption{RAG algorithm configuration space from the database perspective.}
\vspace{-1em}

\scalebox{0.82}{ %
\begin{tabular}{p{0.15\linewidth} p{0.2\linewidth} p{0.8\linewidth}}
\toprule
\textbf{Category} & \textbf{Configuration} & \textbf{Explanation} \\
\midrule



\multirow{8}{*}{\centering\textbf{Database}} 
& Number of documents & Ranges from a single document to enterprise-scale corpora to the entire internet. \\
& Document chunking & Fixed or variable chunk sizes, and whether there are token overlaps between chunks. \\
& Retrieval method & Vector search, keyword-based matching (e.g., BM25), hybrid search, etc. \\
& Embedding model & The model used to encode each passage into a database vector (if vector search is used). \\
& Top-$K$ & Number of the most relevant documents to return per query. \\
& Retrieval quality & Quality requirement measured by recall, precision, or other metrics. \\
& Query rewriting & Whether and how to rewrite the original query (e.g., rephrasing or decomposing into sub-questions). \\
& Result reranking & Whether to rerank retrieved results, and which model is used for reranking. \\
\midrule

\multirow{3}{*}{\shortstack[l]{\textbf{Model-Database} \\ \textbf{Interaction}}}
& Retrieval frequency & One retrieval per generation or multiple retrievals (either at a fixed interval or triggered dynamically). \\
& Integration method &  Retrieved content can be integrated either through the prompt or using pre-computed KV cache.  \\
& Relevance filtering & Whether the model is guided to ignore irrelevant retrieved content during generation. \\

\bottomrule
\end{tabular}
} 
\label{tab:rag-config-space}
\end{table*}

%% file: tables/system_space.tex
\begin{table*}[t]
\centering
\caption{RAG system design space.}
\vspace{-1em}

\scalebox{0.82}{ %
\begin{tabular}{p{0.15\linewidth} p{0.2\linewidth} p{0.8\linewidth}}
\toprule
\textbf{Category} & \textbf{Configuration} & \textbf{Explanation} \\
\midrule

\multirow{3}{*}{\centering\textbf{Hardware}} 
& Retrieval hardware & CPU-based (in-memory or on-disk) or accelerator-based (e.g., GPU, FPGA). \\
& Inference hardware & Hardware used to serve the main LLM and auxiliary models (e.g., rewriter, reranker), such as GPU or TPU. \\
& Resource mapping &  Allocation of compute resources and placement of RAG components across hardware. \\


\midrule

\multirow{4}{*}{\centering\textbf{Software}} 
& Inference library & LLM serving frameworks such as vLLM, SGLang, etc. \\
& Database & Relational or vector databases (e.g., PostgreSQL, Milvus), or specialized search libraries (e.g., FAISS or ScaNN). \\
& Index & Type of index used, such as inverted-file indexes or graph-based indexes, as well as the index parameters. \\
& Quantization & Whether quantization is applied to database vectors (e.g., product quantization or scalar quantization). \\

\midrule

\multirow{2}{*}{\shortstack[l]{\textbf{Cross-Stack} \\ \textbf{Optimizations}}}
& KV-cache reuse & Store and reuse the KV cache of frequently accessed documents, at the cost of losing cross-document attention. \\
& Speculative retrieval & Hiding retrieval latency via approximate document prefetching at the cost of retrieval quality. \\

\bottomrule
\end{tabular}
} 
\label{tab:rag-system-design-space}
\end{table*}

%% file: sections_submission/goal.tex
\section{Goal: Exploring the Pareto Frontier between RAG Quality and Performance}
\label{sec:goal}

Given the complexity of RAG algorithms and systems, existing research has largely focused on either improving generation quality from the algorithmic side~\cite{borgeaud2022improving, asai2023self, ma2023query, glass2022re2g}, or optimizing performance by building efficient systems~\cite{rago:isca:2025, jiang2023chameleon, hu2025hedrarag, hu2025patchwork}.

\textbf{Problem Formulation.} However, real-world deployments often aim to offer users high-quality RAG solutions at different cost levels. 
This raises a fundamental question: how can we efficiently navigate the RAG algorithm and system configuration space to identify the quality–performance Pareto frontier?
%

More specifically, the problem can be formulated as follows: \textit{given a set of available models $M$, a dataset $D$ used to construct the database, an RAG application $A$, and a collection of hardware resources $H$ (e.g., CPUs, GPUs), identify the (near-)Pareto frontier between quality $q$ and performance $p$, while minimizing the number of trials over RAG algorithm and system configurations.}

%
%

\textbf{Challenges in Quality-Performance Co-Optimization.}
Given the large algorithmic and system design spaces, the joint optimization problem spans their Cartesian product, making exhaustive evaluation intractable.
One might suggest assembling a team of experts who understand both the algorithmic and systems aspects of RAG to manually search for the quality-performance Frontier for each application.  

However, expert-driven tuning has two key limitations:
First, it is difficult to find individuals or teams with deep expertise across models, databases, inference frameworks, and hardware, due to the large knowledge gap between these domains.  

Second, \textit{expert knowledge is often not transferable across deployment settings or over time.}
This is because RAG configurations that work well in one setting (with specific application and software/hardware infrastructure) may become suboptimal as the surrounding ecosystem changes.
For example, two organizations may face different constraints across the stack: one may use GPT while the other uses Gemini; one may rely on vector databases, the other on specialized vector search libraries; and their hardware backends may differ (e.g., GPUs, TPUs, etc.). 
These differences can lead to vastly different RAG algorithm–system combinations on the Pareto frontier.  
More importantly, even within the same organization, the RAG ecosystem is continuously evolving: new, more capable models are regularly released; new retrieval algorithms introduce different quality–performance trade-offs; software stacks for LLM inference and vector search improve rapidly; and hardware capabilities for both retrieval and generation shift over time.

%% file: sections_submission/solution.tex
\section{Proposed Solution: \ours}
\label{sec:solution}

\textbf{Overview.}
In this paper, we propose \ours, a three-pillar blueprint for systematically identifying the quality–performance Pareto frontier in RAG serving by navigating the large joint algorithm–system design space.  
The key idea is to explore the RAG space iteratively from the algorithmic side rather than the systems side.
This is because each algorithm configuration corresponds to a specific generation quality and a set of optimal system configurations (with trade-offs across latency, throughput, etc.), while a system configuration can typically support almost arbitrary RAG algorithms if sufficient hardware resources are allocated.

More specifically, the plan exploration process is supported by the following three pillars.
The first pillar, \ir, is a RAG workload abstraction that decouples algorithm-level quality optimization from system-level performance optimization~(\S\ref{sec:solution:ir}).  
The second pillar, \perf, is a cost model that estimates or returns the true system performance for a given RAG algorithm configuration described using \ir~(\S\ref{sec:solution:perf}).  
The third pillar, \plan, is a plan exploration algorithm that leverages \perf to guide the search over RAG algorithm configurations, aiming to find the quality–performance Pareto frontier given the user-defined application and available hardware resources~(\S\ref{sec:solution:plan}).

%

\subsection{RAG-IR as the Workload Abstraction}
\label{sec:solution:ir}

\begin{figure}[t]
  \centering
  \includegraphics[width=0.8\linewidth]{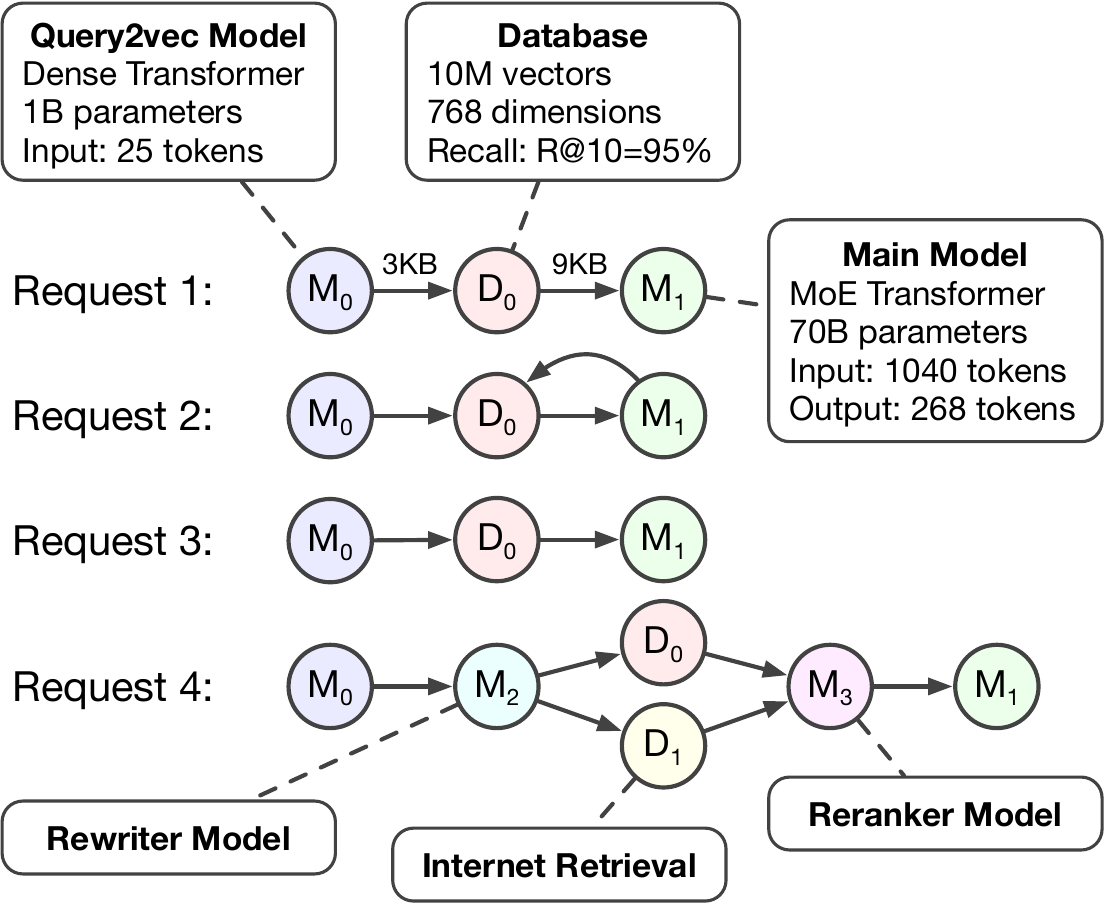}
  \vspace{-1em}
  \caption{\ir is a collection of dataflow graphs. Each node in the graph is either a database or an inference component. }
  \vspace{-1.5em}
  \label{fig:rag-ir}
\end{figure}

\textbf{The Need for Proper Abstraction.} Given the significant knowledge gap between the algorithmic and system spaces and the complexity inherent in each, we envision that quality–performance co-optimization requires an abstraction that cleanly separates the quality and performance dimensions.  
In particular, the abstraction should capture the minimal set of information about an algorithm configuration needed to estimate system performance.  
RAGO~\cite{rago:isca:2025} introduced RAG-Schema, which models RAG pipelines by describing the connections between model and database components, under the assumption of homogeneous workloads (e.g., fixed sequence lengths and pipeline paths) across all requests.  
However, modern RAG workloads are increasingly agentic and heterogeneous: different requests may trigger different execution paths --- for example, one query may invoke a query rewriter or iterative retrieval, while another may not. 
As a result, capturing per-request workload variability is critical to building a meaningful RAG workload abstraction.

\input{tables/rag-ir}

\textbf{Abstraction Overview.}  
We propose \ir, a collection of flexible dataflow graphs defined per request, designed to capture the diverse and agentic nature of modern RAG workloads, as shown in Figure~\ref{fig:rag-ir}.  
Each node represents either a model or a retrieval component, annotated with performance-relevant attributes, while an edge represents execution paths.  
As summarized in Table~\ref{tab:rag-ir}, RAG-IR captures only the minimal and essential information required from the algorithm configuration space (as described in Table~\ref{tab:rag-config-space}), omitting properties that affect only quality --- for example, specific model versions or training data.  
Additionally, RAG-IR annotates workload-specific characteristics such as input/output sequence lengths per inference component as well as inter-component data movement to support fine-grained performance estimation.

\textbf{Exposing Algorithm–System Co-Design Solutions.}
As discussed in \S\ref{sec:complexity:system}, there are emerging algorithm–system co-design techniques that improve system performance but at the potential cost of generation quality.
Some popular examples include KV-cache reuse for frequently accessed documents~\cite{jin2024ragcache, yao2024cacheblend} and speculative retrieval~\cite{jiang2025piperag, zhang2024accelerating}.
We expose such techniques as optional attributes in Table~\ref{tab:rag-ir}; when enabled, it is the responsibility of algorithm designers to re-evaluate end-to-end generation quality.

\vspace{-.5em}
\begin{tcolorbox}[
    enhanced,
    arc=2mm, 
    outer arc=2mm, 
    boxrule=0.8pt, 
    colframe=black, 
    colback=white, 
    boxsep=0pt, 
    drop shadow southeast, 
]

\textbf{Research Gap 1: Should index configuration be part of the RAG workload abstraction?}
\end{tcolorbox}
\vspace{-.5em}

%
An open question in the design of \ir is whether database index configurations should be explicitly provided from the algorithmic side.  
On one hand, RAG algorithm researchers often select the index type and tuning parameters as part of the algorithm design, since generation quality is directly evaluated based on the retrieved results produced by that index.  
However, the chosen index may not yield optimal system performance. 
Thus, an alternative approach is to expose only a quality requirement (e.g., 95\% recall) in \ir, leaving system designers free to select and optimize the index configuration to meet this target.  
The challenge with this approach is that even if an alternative index satisfies the same recall threshold, it may return a different set of documents, potentially affecting the end-to-end generation quality.  
As a result, we treat the database index configuration as an optional input to the abstraction in the current version of \ir, as shown in Table~\ref{tab:rag-ir}: providing explicit index parameters guarantees generation quality but may sacrifice performance, while delegating index selection to the system side enables performance optimization at the potential cost of quality variation.
We believe future research is needed to systematically evaluate such trade-offs.

\subsection{\perf for Performance Prediction}
\label{sec:solution:perf}


\textbf{Objectives of the Cost Model.}
The second pillar of \ours is \perf, a cost model that estimates or returns the exact system performance for a given RAG algorithm configuration.
Given an algorithm described by \ir and a set of available software and hardware resources, \perf aims to predict system performance in terms of Time-To-First-Token (TTFT) latency, Time-Per-Output-Token (TPOT) latency, Requests Per Second (RPS), Requests Per Dollar, and other relevant metrics.
Since there are often trade-offs between these metrics, the cost model should ideally expose the Pareto frontier across them.
Optionally, \perf may also suggest the optimal system configurations on the performance Pareto frontier, such as how to allocate hardware resources across RAG components or how to partition models and databases across multiple servers or accelerators.
There are several approaches to constructing a cost model:

\textbf{Analytical Cost Model.}
The first approach is to use an analytical model.
One example is the roofline model, where the total execution time is estimated as the maximum of the compute and memory access time, computed based on the total number of FLOPs and memory accesses divided by the hardware's peak compute throughput and memory bandwidth.
More advanced analytical models~\cite{rago:isca:2025, bambhaniya2024demystifying} can offer higher accuracy by incorporating detailed operator-level cost breakdowns, replacing peak bandwidth or FLOP/s estimates with calibrated lookup tables, evaluating various sharding strategies for models and databases, and accounting for data movement across machines.
Overall, analytical models provide an interpretable way to predict system performance, though improving their accuracy often requires calibration on the target hardware and software stack (e.g., benchmarking achievable peak performance under various conditions).

\textbf{ML-based Cost Model.}
A second approach to performance prediction is using machine learning.
These models can be trained on performance data collected either from analytical models or from real measurements on hardware and software stacks.
Compared to analytical models, ML-based approaches may achieve higher accuracy when trained on real-system data, but they require extensive data collection and often sacrifice interpretability.
%
%

\textbf{Real Profiling.} 
A third approach is to profile real systems so as to measure the ground-truth performance.
It is most suitable when the system design space is relatively small --- for example, if there is only one GPU and one CPU server, there is no need to consider partitioning models or databases. 
However, for larger workloads spanning hundreds of GPUs or CPU servers, the number of possible system configurations --- such as sharding and parallelization strategies --- can be huge, making exhaustive profiling highly time-consuming.

\vspace{-.5em}
\begin{tcolorbox}[
    enhanced,
    arc=2mm, 
    outer arc=2mm, 
    boxrule=0.8pt, 
    colframe=black, 
    colback=white, 
    boxsep=0pt, 
    drop shadow southeast, 
]

\textbf{Research Gap 2: Accurate cost modeling for retrieval.}
\end{tcolorbox}
\vspace{-.5em}

%
While several cost models have been developed for LLM inference~\cite{rago:isca:2025, bambhaniya2024demystifying, cao2025amali}, accurate performance modeling for retrievals remains largely underexplored.
Existing retrieval performance models~\cite{jiang2023co,rago:isca:2025} only support the IVF index~\cite{PQ}, and they either target specific hardware platforms~\cite{jiang2023co} or assume simple data partitioning strategies for distributed search~\cite{rago:isca:2025}.
Further research is clearly needed to develop accurate and general cost models that support a variety of index types (e.g., IVF, graph-based, LSH), distributed retrieval (with diverse sharding strategies and load-balancing scenarios), and different hardware backends (e.g., CPUs, GPUs, FPGAs, in-memory or on-disk systems).

\subsection{\plan for Configuration Space Navigation}
\label{sec:solution:plan}

\textbf{Objectives of the Plan Explorer.}
Given the vast space of possible algorithmic and system configurations for RAG serving, the goal of \plan is to efficiently identify algorithm–system combinations --- each defined as a RAG serving \textit{plan} --- that lie on the (near-) Pareto frontier between quality and performance, while minimizing the number of exploration iterations.
This exploration can be performed under various constraints, such as a set of available models $M$, a given retrieval dataset $D$, a specific RAG application $A$, and a pool of hardware resources $H$ (e.g., CPUs, GPUs).

\input{algorithms/plan_exploration}



\textbf{A General Plan Exploration Algorithm.}
In each iteration, \plan selects an algorithm configuration, evaluates its generation quality, and estimates the corresponding system performance.
If the result lies on the Pareto frontier, it is added to the result set.
Based on past evaluations, \plan then determines the next algorithm configuration to explore. 

\begin{tcolorbox}[
    enhanced,
    arc=2mm, 
    outer arc=2mm, 
    boxrule=0.8pt, 
    colframe=black, 
    colback=white, 
    boxsep=0pt, 
    drop shadow southeast, 
]

\textbf{Research Gap 3: Predicting the next database configuration to evaluate.}
\end{tcolorbox}
%
An open question here is how to choose the next configuration to evaluate (line 8 in Algorithm~\ref{alg:pareto-search}).
In highly constrained settings --- e.g., with only a few available models or limited hardware resources --- simple approaches like grid search might suffice.
However, in most real-world scenarios, the configuration space is large and complex, requiring alternative approaches such as genetic algorithms or reinforcement learning to efficiently explore the space.
For learning-based approaches in particular, the core challenge lies in how to predict or select the next promising configuration to evaluate.
Different algorithm configuration changes come with different costs: some changes are relatively easy to make, such as swapping the model or adjusting the Top-K value; however, it is costly to rebuild the database --- for instance, switching to a different index type, modifying the chunking strategy, or using a different embedding model to re-encode the database vectors.
We believe efficient plan exploration is an important direction that warrants further research.


%% file: tables/rag-ir.tex
\begin{table}[t]
\centering
\caption{RAG-IR is a dataflow graph composed of nodes and edges, each annotated with the following attributes.}

\vspace{-1em}
\small
\begin{tabular}{p{0.33\linewidth} p{0.56\linewidth}}
\toprule
\textbf{Category} & \textbf{Attributes} \\
\midrule

\multirow{5}{*}{Node (Database)} 
& Number of rows \\
& Vector dimensionality \\
& Top-$K$ \\
& Retrieval quality \\
& (Optional) Index configurations  \\
& (Optional) Enable speculative retrieval  \\

\midrule

\multirow{5}{*}{Node (Model)} 
& Model architecture \\
& Model size \\
& Input sequence length \\
& Output sequence length \\
& (Optional) Enable KV-cache reuse  \\

\midrule

\multirow{1}{*}{Edge (Communication)} 
& Amount of data transfer \\

\bottomrule
\end{tabular}
\label{tab:rag-ir}
\vspace{-1em}
\end{table}

%% file: algorithms/plan_exploration.tex
\begin{algorithm}[t]
\caption{Example Plan Exploration Algorithm \plan}
\label{alg:pareto-search}
\KwIn{Maximum iterations $N$, available models $M$, dataset $D$, RAG application $A$, hardware resources $H$.}
\KwOut{Set of Pareto-optimal configurations $\mathcal{C}$}

Initialize Pareto configuration list $\mathcal{C} \gets \emptyset$ \; 
Initialize algorithm configuration $a_0 \gets \texttt{Init}(M, D, A)$ \;

\For{$i = 1$ \KwTo $N$}{

    Evaluate quality $q_i \gets \texttt{EvalQual}(a_i, A)$ \;
    Predict performance $\hat{p}_i \gets \texttt{\perf}(\texttt{\ir}(a_i), H)$ \;

    \If{$(q_i, \hat{p}_i)$ is on the Pareto frontier}{
        $\mathcal{C} \gets \mathcal{C}\texttt{.Update}\{(a_i, q_i, \hat{p}_i)\}$ \;
    }
    
    Configuration: $a_{i+1} \gets \texttt{\plan.GetNext}(a_{0:i}, q_{0:i}, \hat{p}_{0:i})$ \;
}

\Return{$\mathcal{C}$}
\end{algorithm}

%% file: sections_submission/conclusion.tex
\section{Conclusion and Future Work}

We present \ours, the first blueprint for systematically optimizing RAG serving from both performance and quality perspectives.
We are working on building a prototype of \ours using a limited set of models, system components, and hardware platforms to demonstrate the effectiveness of the proposed combination between abstraction \ir, cost model \perf, and plan explorer \plan.
We believe \ours will serve as the foundation for future research on quality–performance co-optimization in RAG serving systems.